\documentclass[conference]{IEEEtran}

\usepackage{amsmath,amssymb,amsfonts}
\usepackage{graphicx}
\usepackage{textcomp}
\usepackage[table]{xcolor}
\usepackage[utf8]{luainputenc}
\usepackage[british]{babel}
\usepackage[acronym, nomain]{glossaries}
\usepackage{marginnote}
\usepackage{tikz}
\usetikzlibrary{arrows.meta, positioning, shapes.misc, calc, patterns}
\usepackage{tikz-dimline} 
\usepackage{float}
\usepackage{tabularx}
\usepackage{booktabs}
\usepackage[backend=biber,style=ieee, mincitenames=1, maxcitenames=2, url=true, isbn=true, natbib]{biblatex}
\usepackage{usebib}
\usepackage[hidelinks]{hyperref}
\usepackage[capitalise]{cleveref}
\usepackage{marginnote}
\usepackage{csquotes}
\usepackage[utf8]{inputenc}
\usepackage[T1]{fontenc}
\usepackage{textcomp}

\def\BibTeX{{\rm B\kern-.05em{\sc i\kern-.025em b}\kern-.08em
		T\kern-.1667em\lower.7ex\hbox{E}\kern-.125emX}}

\AtEveryBibitem{%
	\ifentrytype{article}{%
		\clearfield{url}%
		\clearfield{urldate}%
	}{}%
	\ifentrytype{inproceedings}{%
		\clearfield{url}%
		\clearfield{urldate}%
	}{}%
}
\DeclareFieldFormat{urldate}{%
	\ifentrytype{article}{}{
		\ifentrytype{inproceedings}{}{
			\mkbibparens{\bibstring{urlseen}\space#1}
		}%
	}%
}

\usetikzlibrary{patterns}

\bibinput{references}

\newbibfield{author}
\newbibfield{title}
\newbibfield{year}

\newcommand{\printyear}[1]{\usebibentry{#1}{year}}
\newcommand{\fullc}{\tikz\fill[black] (0,0) circle (0.1cm);}
\newcommand{\quarc}{
	\tikz{
		\draw[black] (0,0) circle (0.1cm);
		\fill[black] (0,0) -- (0.1cm,0) arc [start angle=0, end angle=-90, radius=0.1cm] -- cycle;}}
\newcommand{\halfc}{\tikz{
		\draw[black] (0,0) circle (0.1cm);
		\fill[black] (0,0) -- (-0.1cm,0) arc [start angle=180, end angle=360, radius=0.1cm] -- cycle;}} 
\newcommand{\emptyc}{\tikz\draw[black] (0,0) circle (0.1cm);}

\usepackage{caption} 
\usepackage{algorithm}
\usepackage{algpseudocode}
\usepackage{pgf}

\newacronym{i4aas}{I4AAS}{Industry 4.0 Asset Administration Shell}
\newacronym{hmi}{HMI}{Human-Machine Interface}
\newacronym{opcua}{OPC UA}{Open Platform Communications Unified Architecture}
\newacronym{jrc}{JRC}{Josef Ressel Centre}
\newacronym{dmz}{DMZ}{Demilitarised Zone}
\newacronym{isia}{ISIA}{Intelligent and Secure Industrial Automation}
\newacronym{imm}{IMM}{Injection Moulding Machine}
\newacronym{opc}{OPC}{Open Platform Communications}
\newacronym{gds}{GDS}{Global Discovery Service}
\newacronym{pki}{PKI}{Public Key Infrastructure}
\newacronym{plc}{PLC}{Programmable Logic Controller}
\newacronym{scada}{SCADA}{Supervisory Control and Data Acquisition}
\newacronym{mqtt}{MQTT}{Message Queuing Telemetry Transport}
\newacronym{https}{HTTPS}{Hypertext Transfer Protocol Secure}
\newacronym{uri}{URI}{Uniform Resource Identifier}
\newacronym{it}{IT}{Information Technology}
\newacronym{ot}{OT}{Operational Technology}
\newacronym{dos}{DoS}{Denial of Service}
\newacronym{rtu}{RTU}{Remote Terminal Unit}
\newacronym{ids}{IDS}{Intrusion Detection System}

\crefname{figure}{Figure}{Figures}
\Crefname{figure}{Figure}{Figures}
\crefname{table}{Table}{Tables}
\Crefname{table}{Table}{Tables}

\begin{document}
\title{Secure Data Bridging in Industry 4.0: An OPC UA Aggregation Approach for Including Insecure Legacy Systems}
\author{
	\IEEEauthorblockN{Dalibor Sain\IEEEauthorrefmark{1}, 
						Thomas Rosenstatter\IEEEauthorrefmark{1}, 
						Olaf Saßnick\IEEEauthorrefmark{1},
						Christian Schäfer\IEEEauthorrefmark{2},
						Stefan Huber\IEEEauthorrefmark{1}	}
	\IEEEauthorblockA{\IEEEauthorrefmark{1} Josef Ressel Centre for Intelligent and Secure Industrial Automation\\
		Salzburg University of Applied Sciences, Austria\\
		\{dalibor.sain, thomas.rosenstatter, olaf.sassnick, stefan.huber\}@fh-salzburg.ac.at}
	\IEEEauthorblockA{\IEEEauthorrefmark{2} B\&R Industrial Automation GmbH, Salzburg, Austria\\
		christian.schaefer@br-automation.com}
}

\maketitle

\thispagestyle{plain}
\pagestyle{plain}

\begin{abstract}
	The increased connectivity of industrial networks has led to a surge in cyberattacks, emphasizing the need for cybersecurity measures tailored to the specific requirements of industrial systems. Modern Industry 4.0 technologies, such as OPC UA, offer enhanced resilience against these threats. However, widespread adoption remains limited due to long installation times, proprietary technology, restricted flexibility, and formal process requirements (e.g. safety certifications). Consequently, many systems do not yet implement these technologies, or only partially. This leads to the challenge of dealing with so-called brownfield systems, which are often placed in isolated security zones to mitigate risks. However, the need for data exchange between secure and insecure zones persists.
	
	This paper reviews existing solutions to address this challenge by analysing their approaches, advantages, and limitations. Building on these insights, we identify three key concepts, evaluate their suitability and compatibility, and ultimately introduce the SigmaServer, a novel TCP-level aggregation method. The developed proof-of-principle implementation is evaluated in an operational technology (OT) testbed, demonstrating its applicability and effectiveness in bridging secure and insecure zones.
\end{abstract}

\newcommand{\sigmaserv}{SigmaServer}
\begin{IEEEkeywords}
    Industry 4.0, OPC UA, Security, Aggregation Server, Brownfield, Retrofit
\end{IEEEkeywords}

\section{Introduction}
\label{sec:introduction}

\subsection{Motivation}
By applying the design principles of Industry 4.0, modern industrial plants are evolving into highly interconnected \gls{ot} systems, often also linked to the worldwide Internet.
While providing benefits such as higher efficiency and greater flexibility, the interconnectedness also increases the overall attack surface, resulting in higher cybersecurity risks~\cite{Briggs2019,Miller2021OTattacks}.

To address these risks, initiatives like the IEC 62443~\cite{iec62443-3-2} put focus on cybersecurity for \gls{ot},
proposing the use of zones and conduits to strengthen the security of the industrial network.
As such, a network is separated into multiple zones with different security requirements, following a defense-in-depth concept.
Zones are connected by conduits, which are typically firewalls or one-way gateways (data diodes), restricting unauthorised traffic. 

A remaining challenge however is the long operational lifespan of \gls{ot} equipment. 
While IT components like desktop and server systems are typically used for 3-5 years,
a production machine represents a far larger investment and is commonly used for 10-15 years and longer~\cite{nist-sp800-82r3}. 
The hardware and software embedded in these machines cannot easily be kept up to date and therefore at some point
in their lifespan no longer support up-to-date security standards (such as secure OPC UA communication).
They become legacy devices, providing only reduced or no security measures.
As such they pose a risk to the entire industrial network, significantly weakening the overall security.
In addition, some protocols may still be popular in the industry, yet do not provide any or sufficient security measures. 

In this paper, we use the term {\it legacy device} to refer to devices that are {\it not able to fulfill the security requirements} of modern industrial networks -- including also insecure protocols such as Modbus.

To mitigate the insecurity of devices, they can be placed in fully isolated zones.
The isolation prevents attacks but at the same time removes the benefits of connectivity, such as precise synchronization and process optimisation.
Therefore, to a certain extent, insecure devices still must be able to communicate beyond their isolated zone, for instance, with \gls{scada} systems. 

This can be realized with Middleware solutions, bridging the gap between isolated zones containing legacy devices and secure zones with systems supporting modern security features.
Such solutions must not only be able to translate between different protocols, but also must represent legacy devices beyond their isolated zone with up-to-date security features.
Especially the latter requirement is currently not well addressed by existing solutions, as the primary focus is a working, fast and reliable translation between protocols.
Therefore, in the following work, our focus is on how to securely integrate legacy brownfield systems in modern secure  industrial networks.

\subsection{Contribution}

Our contribution towards securing industrial legacy systems via aggregation in modern production environments is threefold:
\begin{itemize}
	\item We provide a structured review of published middleware aggregation solutions and analyse them with regard to their security capabilities. 
	\item As none of the published solutions are found to focus on security, we define a common threat model and introduce three general architectural concepts for embedding insecure legacy systems in a modern, secure industrial network aligned with the Purdue model~\cite{Purdue1989Williams}.
	\item We implement the most practical of these concepts as a proof-of-principle, named \sigmaserv{}, and evaluate its performance characteristics, such as latency and system load, demonstrating that retrofitting can be achieved with acceptable technical effort and low system overhead.
\end{itemize}
The evaluation shows that our proposed \sigmaserv{} achieves an end-to-end latency below 2.6 ms, while its internal processing delay remains in the low microsecond range with an average of 21.15 $\mu$s.
\sigmaserv~requires significantly less RAM usage (6–19 MiB) compared with the \gls{opc} Foundation Console Aggregation Server\footnote{\url{https://github.com/OPCFoundation/UA-.NETStandard-Samples/tree/master/Workshop/Aggregation/ConsoleAggregationServer/}} (105–115 MiB). Although CPU usage is slightly higher, it remains in the low single-digit range  with 0.75–3.16\%.
For reproducibility and further research, our implementation is also made publicly available on GitHub\footnote{\url{https://github.com/JRC-ISIA/opc-ua-sigmaserver/}}.

\subsection{Organisation of the Paper}
The rest of the paper is organised as follows. 
\Cref{sec:background} provides background information on industrial networks and zones, and an overview of OPC UA.
Furthermore, this section introduces the reference architecture which is followed throughout this paper. 
\Cref{sec:methodology} presents the methodology, which is based on two research questions.
\Cref{sec:stateoftheart} reviews the state of the art and analyses existing solutions with regard to their security capabilities. 
\Cref{sec:concepts} introduces the threat model and three architectural concepts which are evaluated based on their suitability and compatibility with industrial networks. 
The design of the proposed solution, \sigmaserv{}, is further detailed in \Cref{sec:sigmaserv}.
\Cref{sec:experimental:setup} and~\ref{sec:results:discussion} describe the setup of the evaluation experiments and the results respectively.
Finally, \cref{sec:conclusion} concludes the paper and provides an outlook on future work.

\section{Background}
\label{sec:background}

\subsection{Industrial Networks}\label{subsec:industrial-networks}

Industrial networks form the technological backbone of modern automation environments.
They enable the interaction between sensors, actuators, controllers, and supervisory systems. 

At the core of these networks are \glspl{plc}, which are located on the field level and execute deterministic control tasks within strict timing constraints. 
Such real-time operation ensures that industrial processes, which range from manufacturing lines to energy grids, function deterministically and reliably~\cite{knapp2014industrial}.

In contrast to conventional \gls{it} infrastructure, \glspl{plc} often rely on specialised industrial communication technologies beyond TCP/IP over Ethernet. 
They often additionally use fieldbuses or industrial Ethernet-based technology (e.g., POWERLINK, PROFINET, EtherCAT) that better satisfy real-time and reliability requirements essential to process control.

One of these protocols is Modbus which is both open and widely adopted in the industrial domain~\cite{URREA201627}
This widespread adoption can be attributed to its inherent simplicity allowing interoperability among devices.
Originally, Modbus was developed for serial communication through Modbus RTU and Modbus ASCII. The Modbus protocol has since evolved to include Ethernet-based communication via Modbus TCP as part of IEC~61158~\cite{IEC61158-6}, thereby accommodating a broad spectrum of industrial applications and use cases.
Given that the Modbus variant including security features based on TLS, known as Modbus/TCP Security, was not introduced until 2018, its adoption remains limited, which is understandable considering the extended operational lifetimes typical of industrial control systems. Moreover, Modbus/TCP Security is only defined by the Modbus Organisation -- it is not part of IEC~61158.

Above the level of device control (\glspl{plc}, actuators and sensors), \gls{scada} systems offer visualisation, data logging, and high-level management of processes distributed across multiple sites. 
\gls{scada} systems integrate with PLCs, \glspl{rtu}, and \glspl{hmi} to provide operators with a comprehensive view of system performance and state. 
Together, these components create a hierarchical architecture where local, time-critical control functions operate in an environment alongside with centralised monitoring and decision-making~\cite{knapp2014industrial}.
\subsection{Industry 4.0 and OPC UA}

The term \textit{Industry~4.0}, the fourth industrial revolution, is characterised by the convergence of physical, industrial systems with traditional \gls{it} systems. 
The goal is to create production systems, which are capable of self-optimisation and autonomous decision-making~\cite{lasi2014}.

For that purpose, flexibility, adaptability, transparency and interoperability across heterogeneous systems must be met~\cite{schleipen2016}. 
\citeauthor{schleipen2016}~\cite{schleipen2016} mentions, that such a compliant system must therefore be able to communicate via open, standardised \gls{it} networks, be self-describing and protective about its own information, and also be able to configure and optimise themselves. 
With the so-called RAMI~4.0 model~\cite{ZVEI2015RAMI40}, a reference architecture for Industry 4.0 systems was developed, which specifies the needed semantic technologies.

Furthermore, a strategy paper by the German Federal Government~\cite{umsetzungstraetegie-2015}, identifies \gls{opcua} as a foundational technology for Industry~4.0.
This is also strengthened in a report discussing the criteria for Industry~4.0 by ZVEI~\cite{leitfaden_2018}. 
\gls{opcua} supports security features such as access control, and provides platform independence through its information model and decoupling from the database model making it a well-suited solution for industrial communication~\cite{schleipen2016}.

\textbf{\gls{opcua}}, the successor of the OPC standard, has become a widely adopted standard for industrial communication. It is standardised in IEC~62541 and maintained by the \gls{opc} Foundation. 
The platform-independent, secure-by-design and service-oriented architecture, enables communication for industrial systems~\cite{opcfoundation2022-01}, either in a client-server fashion, or using a publish-subscriber model.

\gls{opcua} defines not only a communication protocol but also introduces information models to structure variables, methods, and events~\cite{crespi2019}.
The information model consists of nodes (objects, variables, methods, events, etc.) and references between them, organised within the address space of an \gls{opcua} server.
\gls{opcua} servers at configured endpoints allow the information model’s variables to be read or written and its methods to be invoked.
Each \gls{opcua} node can be addressed using its \gls{uri} and a unique identifier~\cite{opcfoundation2022-01}.

Concerning its security, the specification defines a set of security features, which help postulate a secure-by-design architecture. 
These features include secure session management, transport security, user authentication and access control.

\textbf{OPC UA aggregating servers} connect multiple \gls{opcua} servers and provide a single, centralised, aggregated address space as.
\citeauthor{grossmann2014}~\cite{grossmann2014} propose a reference architecture for such a server, which contains components for providing the aggregated address space, fetching underlying data, discovery and security management for access control.

\begin{figure}
	\centering
	\includegraphics[width=.85\linewidth]{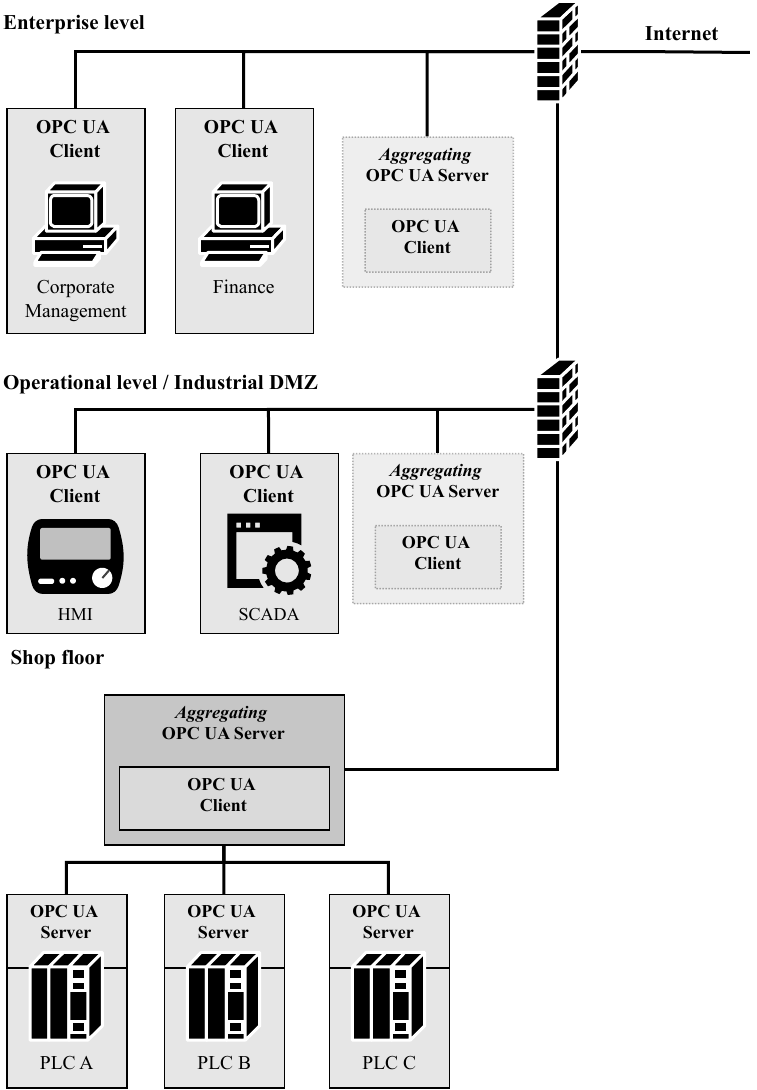}
	\caption{Deployment scenarios of an \gls{opcua} aggregating server providing a single, aggregated address space~\cite{grossmann2014,Schleipen2019}.}
	\label{fig:opcua-aggregation-server}
\end{figure}

Figure~\ref{fig:opcua-aggregation-server} presents various deployment scenarios of an \gls{opcua} aggregating server.  
It illustrates an industrial network architecture inspired by the Purdue model~\cite{Purdue1989Williams} and the recommendations in IEC~62443~\cite{iec62443-3-2}. 
This network is separated into three zones, which are connected through firewalls: 
(i)~the shop floor represents the field level (see Section~\ref{subsec:industrial-networks}) containing the \glspl{plc}, (ii)~the operational level contains monitoring and high-level control of the process via a \gls{scada} system and an \gls{hmi} for user interaction, and (iii)~the highest level, the enterprise zone, contains enterprise systems like financing and corporate management.
In Figure~\ref{fig:opcua-aggregation-server}, three possible deployment scenarios are highlighted, with the aggregating server located either on the shop floor or at one of the higher levels, depending on the specific use case.

The primary advantage of deploying aggregating servers in each of the three scenarios is that they serve as a centralised platform for external networks, providing access to information collected from internal \gls{opcua} nodes.
This advantage becomes clear when considering a scenario where many clients access multiple, often the same, servers: managing multiple connections and users with varying privileges, each requiring the enforcement of individual security rules, demands significant resources~\cite{grossmann2014}. 

By using an aggregating server, the number of connections and users is also reduced to one per client, which also simplifies security management. 
Maintaining only a single connection is an additional advantage for resource-constrained devices with real-time tasks,
 such as \glspl{plc}, which may only be capable of handling a limited number of simultaneous connections.

\subsection{Reference Architecture of an Industrial System}\label{subsec:ref-arch}

This section presents a reference architecture to analyse and demonstrate a security-hardened aggregation service that connects legacy devices to an \gls{ot} network.

\begin{figure}
	\centering
	\includegraphics[width=.84	\linewidth]{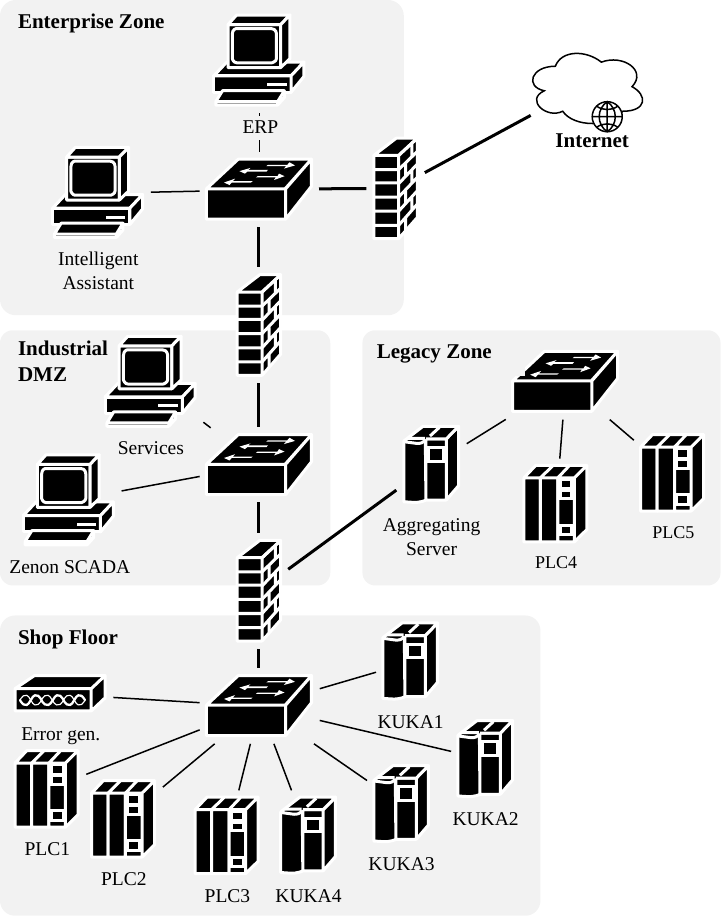}
	\caption{Network setup of the JRC ISIA testbed showing the segmentation into four zones: enterprise zone, industrial \gls{dmz}, legacy zone,  and shop  floor.}
	\label{fig:ref:architecture}
\end{figure}

The reference architecture shown in Figure~\ref{fig:ref:architecture} is based on the \gls{jrc} for \gls{isia} testbed, a testbed for industrial research in which all \gls{ot} devices use the OPC~UA protocol. 
The testbed is structured in accordance with the state of the art, following the IEC~62443~\cite{iec62443-3-2} and the Purdue reference architecture~\cite{Purdue1989Williams}. 
It consists of four security zones that are each connected through a single conduit via firewalls: the \textit{enterprise} zone, the \textit{industrial} \gls{dmz}, the \textit{shop floor} zone, and the \textit{legacy} zone.

The process reflected in this testbed is the production of plastic parts with \glspl{imm}. 
The three \glspl{imm} are each controlled by one \gls{plc} (\textit{PLC1-3}) on the shop floor and a \gls{scada} \gls{hmi} in the operational zone.

All \gls{ot} devices communicate via the \gls{opcua} protocol.
While the industrial computers (\textit{KUKA1--4}) each control a robot arm to move finished parts from the \glspl{imm} and place them on a conveyor belt; a fourth robot arm then puts them on pallets at the line's end.
Although the \glspl{imm} and robot arms are simulated,
the control logic executes identically to a real-world deployment.
The \textit{enterprise} zone comprises a resource-planning tool and an intelligent assistant (part of separate research on intelligent automation),
whereas the \textit{legacy} zone, the focus of this research, connects via a firewall and hosts the proposed \sigmaserv{} server, one \gls{plc} (\textit{PLC4}), and a Raspberry Pi emulating an insecure Modbus temperature controller (\textit{PLC5}).

Taken together, the chosen testbed models a modern industrial manufacturing process using state-of-the-art protocols (\gls{opcua}) to advance research in cybersecurity and AI-driven automation and anomaly detection.

\section{Methodology}
\label{sec:methodology}
The methodological approach used to investigate the secure integration of legacy systems is guided by two research questions:

\begin{enumerate}
	\item[\textbf{RQ.1}] \textit{What security and architectural solutions exist for integrating legacy systems into OPC UA-based industrial automation networks?}
	\item[\textbf{RQ.2}] \textit{Which integration approaches offer the best balance of security and operational feasibility for legacy system inclusion in industrial networks that follow the Purdue model? And how can such an approach be realised?}
\end{enumerate}

To address \textbf{RQ.1}, a structured literature review is conducted 
with the goal of identifying the state of the art regarding published middleware aggregation solutions.
The review aims to provide a broad picture of current approaches for aggregating or retrofitting actuators, sensors, and \glspl{plc}. 
To achieve this, the following search terms are used: \textit{opc~ua, security, aggregation, aggregating, server, brownfield, retrofit}.
The search is conducted via Google Scholar to minimise potential bias in favour of any single publisher~\cite{Wohlin2014a}.
Consequently, the identified publications are classified according to characteristics relevant to security and architecture, as detailed in Section~\ref{sec:stateoftheart}.

With the state of the art established, \textbf{RQ.2} puts focus on underlying architectural concepts for integrating legacy systems into modern industrial networks.
Based on a derived threat model, different architectural concepts for integrating insecure legacy systems into a secure, 
Purdue-compliant network are assessed for security and usability likewise.

Following the identification of the most suitable concept, a proof-of-principle is to be realised in an industrial testbed. 
The proof-of-principle is then to be evaluated using metrics such as latency and system load, to provide insight into the trade-offs between security and operational overhead.
 
\section{State of the Art Solutions}
\label{sec:stateoftheart}
\newcommand{\printbibinfo}[1]{\cite{#1} & \printyear{#1} & \raggedright \citeauthor{#1}}
\begin{table*}[!t]
	\caption{Literature overview on state of the art publications (until January 2026) on aggregating servers and brownfield/retrofitting systems ordered by publication year.}\label{tab:stateoftheart}
	\centering \renewcommand{\arraystretch}{1.25}
	\begin{tabular}{@{}ccp{2cm}ccccccccp{5cm}@{}}
		\toprule
		\textbf{Ref.} 	& \textbf{Year} & \textbf{Authors} &  \textbf{Arch.} & \textbf{IM} &  \textbf{Sec.} & \textbf{T. crit.} & \textbf{Code} & \textbf{Eval.} & \textbf{Leg.} & \textbf{Bf.} & \textbf{Comment}\\
		\multicolumn{3}{l}{\textit{Aggregating Servers}}\\\cmidrule(l){1-3}
		\printbibinfo{fernbach2014} & \emptyc & \fullc & \emptyc & \emptyc & \emptyc & \fullc & \emptyc & \emptyc & Presents a comprehensive approach to connect buildings and industrial automation systems using a cross-domain
		information models. \\
		\printbibinfo{grossmann2014} & \fullc & \fullc & \halfc & \emptyc & \emptyc & \fullc & \emptyc & \emptyc & Present a proof of concept implementation and highlight that aggregating servers bring benefits for security. \\
		\printbibinfo{seilonen2016} & \fullc & \fullc & \emptyc & \emptyc & \emptyc & \fullc & \emptyc & \emptyc & Focuses on the transformation of the information model.\\
		\printbibinfo{Banerjee2017Aggregation} & \fullc & \fullc & \emptyc & \fullc & \emptyc & \emptyc & \emptyc & \emptyc & Uses the prototype of~\cite{grossmann2014} and focuses on unifying data from underlying, heterogeneous services.\\
		\printbibinfo{graube2017} & \emptyc & \fullc & \quarc & \fullc & \emptyc & \emptyc & \halfc & \halfc & Focuses on the challenges of mirroring several OPC UA servers by using co-simulation. \\
		\printbibinfo{wang2018} & \fullc & \fullc & \emptyc & \emptyc & \emptyc & \fullc & \emptyc & \emptyc & Focuses on the cache management.\\
		\printbibinfo{wurger2018} & \emptyc & \fullc & \emptyc & \emptyc & \emptyc & \emptyc & \emptyc & \emptyc & Uses AutomationML to model the
		hierarchical structure and the subsequent mapping to an address space for energy data integration. \\
		\printbibinfo{breunig2019multiprotocol} & \fullc & \fullc & \quarc & \emptyc & \emptyc\textsuperscript{*} & \fullc & \emptyc & \emptyc &  
		Focuses on multi protocol functions.\\
		\printbibinfo{crespi2019} & \fullc & \fullc & \emptyc & \fullc & \emptyc & \emptyc & \emptyc & \emptyc & Security is only mentioned in future work.\\
		\printbibinfo{pena2019} & \emptyc & \emptyc & \emptyc & \fullc & \emptyc & \fullc & \fullc & \fullc & Presents a OPC UA/Modbus gateway solution.  \\		
		\printbibinfo{li2020} & \fullc & \fullc & \emptyc & \fullc & \emptyc & \fullc & \fullc & \emptyc & Focuses on an architecture over TSN.\\
		\printbibinfo{mathias2020} & \halfc & \fullc & \emptyc & \emptyc & \emptyc & \emptyc & \emptyc & \emptyc & Focuses on aggregation dynamics of OPC UA for connecting and accessing multiple database servers. \\
		\printbibinfo{weskamp2021} & \fullc & \fullc & \emptyc & \fullc & \emptyc & \fullc & \emptyc & \emptyc & Uses the architecture from~\cite{grossmann2014}.\\
		\printbibinfo{pu2022} & \fullc & \halfc & \emptyc & \fullc & \emptyc & \fullc & \emptyc & \emptyc & Presents a detailed description of the architecture.\\
		\cite{UnifiedAutomation2025} & 2025 & Unified Automation & \fullc & \fullc & \fullc & \halfc & \emptyc & \emptyc & \fullc & \fullc & Commercial solution of an OPC UA gateway for devices using legacy OPC protocols.\\
		\printbibinfo{knobloch2026} & \fullc & \fullc & \quarc & \halfc & \emptyc & \fullc & \emptyc & \emptyc & Presents hash-based type aggregation and reference injection for instance aggregation.\\
		\multicolumn{3}{l}{\textit{Brownfield and Retrofitting}}\\\cmidrule(l){1-3}
		\printbibinfo{etz2020} & \fullc & \fullc & \quarc & \fullc & \emptyc & \emptyc & \fullc & \fullc & Focuses on implementing an OPC UA gateway to bridge telnet with a legacy device using 'python-opcua'.\\

		\printbibinfo{rupprecht2021} & \fullc & \emptyc & \quarc & \emptyc & \emptyc & \emptyc & \fullc & \emptyc & Evaluates retrofitting legacy PLCs into Industry 4.0 with a hardware gateway.\\
		\printbibinfo{majumder2022} & \fullc & \fullc & \emptyc & \emptyc & \emptyc & \fullc & \fullc & \fullc & Uses PLC4X middleware to integrate legacy systems into Industriy 4.0. \\	
		\printbibinfo{tran2022} & \fullc & \emptyc & \fullc & \emptyc & \emptyc & \fullc & \fullc & \fullc & Reviews retrofitting brownfield systems, highlighting security challenges, threats and solutions.\\
		\printbibinfo{r2024} & \fullc & \halfc & \fullc & \fullc & \emptyc & \fullc & \fullc & \fullc & Presents a three layer architecture for integrating legacy systems with OPC UA. \\
		\printbibinfo{holmes2025} & \fullc & \fullc & \quarc & \emptyc & \emptyc & \emptyc & \fullc & \emptyc & Describes a digital retrofitting framework using OPC UA and a layered RAMI 4.0 architecture. \\	
		\bottomrule\\
	\end{tabular}
	
	{\raggedright {\bf Arch.:} Architecture; {\bf IM:} Information Model; {\bf Sec.:} Security; {\bf T.crit.:} Time Criticality; {\bf Code:} Published Code; {\bf Eval.:} Evaluation; {\bf Leg.:} Legacy devices; {\bf Bf.:}~Brownfield devices; \textsuperscript{*}the git project does not exist anymore.}
\end{table*}

The two topics of \gls{opcua} aggregating servers and retrofitting brownfield systems have already been discussed in depth in the past (see \Cref{subsec:Security in OPC UA Aggregating Server Solutions} and~\ref{subsec:Security in Retrofitting Solutions}),
however, when combining both areas, no research publication could be found, which evaluates aggregating servers as a retrofitting solution.

Table~\ref{tab:stateoftheart} provides a summary of the state of the art as found by the literature review in January 2026.
The evaluated characteristics indicate whether a solution focused~\fullc, considered~\halfc, or did not address~\emptyc{} the topic.
Furthermore, for works that identified security, yet did not focus or consider it, \quarc{} is used.

To clarify the security perspective of the existing work, the analysis is divided into two subsections.
\Cref{subsec:Security in OPC UA Aggregating Server Solutions} examines how security is treated in existing \gls{opcua} aggregating server solutions.
\Cref{subsec:Security in Retrofitting Solutions} reviews security aspects across all identified retrofitting approaches for brownfield systems, 
where insecure protocols and outdated systems introduce different challenges.

\subsection{Security in OPC UA Aggregating Server Solutions}\label{subsec:Security in OPC UA Aggregating Server Solutions}
We identified 15 publications and 1 commercial solution for comparison which focus on aggregating services for \gls{opcua}, which are listed in Table~\ref{tab:stateoftheart}.

Großmann et al.~\cite{grossmann2014} present the basic concept of an \gls{opcua} 
aggregating server and propose a reference architecture for such a server. They
developed two information model extensions (for the aggregation- and the
aggregated server)
and implemented a prototype. Besides a \textit{security manager}
for access control, no further security aspects are discussed.

Banerjee and Großmann~\cite{Banerjee2017Aggregation} make use of the same prototype as in
\cite{grossmann2014} and use it as aggregation layer to unify data from
underlying, heterogeneous services with different information models. Security
is not discussed in this publication.

\citeauthor{crespi2019}~\cite{crespi2019} use an aggregation server as a bridge between cloud
services and the shop floor level (edge devices) and provide a single,
homogeneous information model. They go into more detail regarding how the
aggregated address space is constructed. For future work, they mention the
implementation of security mechanisms.

In a research paper of \citeauthor{weskamp2021}~\cite{weskamp2021} from 2021 an Industry 4.0 compliant \gls{opcua} aggregating server is presented. 
They use the same architecture as in \cite{grossmann2014} as base and map the underlying information models to the metamodel of the \gls{i4aas} companion specification. 
Security is not discussed in this publication.
Version 1.0 of the \gls{opcua} specification for compliance with \gls{i4aas} is defined in OPC~30270~\cite{OPC30270AAS} and was published the same year (2021).

\citeauthor{pena2019} presents a Modbus to \gls{opcua} gateway solution implemented on a Raspberry Pi.
The proposed gateway enables communication between legacy Modbus based devices and modern \gls{opcua} clients but differs from our work by focusing on protocol translation rather than on securely bridging isolated network zones.

Other research, such as \cite{pu2022}, \cite{seilonen2016}, \cite{wang2018},
\cite{graube2017} or \cite{li2020} make also use of aggregating servers, but do
not discuss security aspects any further.

The GitHub project of the OPC Foundation \cite{OPCFoundationGithub} provides open-source code for sample OPC UA applications according to the standard.
It also includes samples for an OPC UA aggregating server.
A commercial solution, the UaGateway, is offered by Unified Automation~\cite{UnifiedAutomation2025}, which provides secure communication between devices using the legacy \gls{opc} protocol and those using \gls{opcua}.
Both approaches differ from our proposed aggregating server solution presented in \textit{C.3} Section~\ref{sec:c3tcplevelaggregation} as we also consider other legacy protocols that are neither \gls{opc} or \gls{opcua}. Furthermore, our proposed \sigmaserv{} provides its own secure \gls{opcua} server per device.

\subsection{Security in Retrofitting Solutions}\label{subsec:Security in Retrofitting Solutions}
Six relevant publications on retrofitting of brownfield systems with \gls{opcua} were identified in the review.

\citeauthor{rupprecht2021}~\cite{rupprecht2021} evaluate multiple retrofitting concepts to integrate
legacy \glspl{plc} in a modern, Industry 4.0 compliant environment and provide
their data to higher-level systems. One of the approaches
places an additional hardware gateway at field level, which connects to the
\gls{plc} via its proprietary protocol and to the intranet via \gls{opcua}.
Security is not discussed in this publication.

In \cite{etz2020}, the authors implement a similar solution: an \gls{opcua}
gateway which bridges a telnet connection of a legacy device to an \gls{opcua}
domain. The implementation is done using the information model from
the manufacturer and the library ``python-opcua''. Security is not discussed in
this publication.

In \cite{majumder2022}, \textit{PLC4X} is used as a middleware to integrate
legacy systems with different proprietary protocols into an Industry 4.0
environment using modern, standardised protocols such as \gls{opcua},
\gls{mqtt}, \gls{https}, etc. However, they conclude, that this tool is unable
to handle the different data semantics and data heterogeneity of underlying
devices and is therefor not suitable as standalone tool.

In \cite{tran2022}, a review on previous work regarding retrofitting brownfield
systems is presented. The authors include typical challenges, with one of them
being security. They mention problems with legacy machines, limited security
functions and security threats and their possible solutions. They also note,
that there are only few studies, which include security aspects. Considering
that the use of legacy systems is still very common, they mention that security
is an increasingly important factor for the future.

In contrast to other solutions, the authors of \cite{lackorzynski2020} make use
of a modified MACsec protocol to secure communication on layer 2. 
This has the advantage that industrial protocols based on Ethernet can be secured on the ISO/OSI layer 2. 
However, using an Ethernet based security mechanism does not address the problem of the continued use of legacy protocols within the industrial network.
Moreover, it requires the non standardised use of the MACsec Ethernet frame.

\citet{busboom2024} present a review on works that generate \gls{opcua} information models in an automated fashion.
The publication includes  information in aggregating multiple \gls{opcua} information models, namespace indices, merging types, etc.

In \cite{r2024}, a three layered architecture is used for integrating various legacy devices into Industry 4.0. This publication focuses on the security aspect, but differs to the solution presented in this work.

\citet{pena2019} demonstrated an OPC UA/Modbus gateway for energy recovery systems but did not address the secure integration of other proprietary protocols. In \cite{holmes2025} proposed a general digital retrofitting framework following a layered RAMI 4.0 architecture, using OPC UA as the standard for modular and distributed communication across heterogeneous legacy machines.

\citet{knobloch2026} propose a hash-based method for type aggregation and an instance aggregation approach using reference injection.
While they validate this mapping database design on a robotic assembly station, their focus remains on vertical integration within secure networks.

In summary, of the 22 publications reviewed, only three focus specifically on security, one considers it to some extent, and six identify a need for further security measures in the future.
The characteristic \textit{Code} (see Table~\ref{tab:stateoftheart}), highlights that none of the identified publication provide their implementation as open-source, thus limiting the reproducibility.

Furthermore, we can summarise our findings as follows:

\begin{itemize}
	\item \gls{opcua} aggregating servers provide an effective solution for integrating multiple underlying \gls{opcua} servers into a unified, aggregated address space.
	\item Retrofitting brownfield systems with \gls{opcua} gateways is a common approach for bridging the gap between legacy systems and modern, Industry 4.0-compliant infrastructures. 
	\item Security in retrofitting brownfield systems is often a secondary concern and receives only limited attention. 
	\item None of the proposed works have published a proof-of-principle, hindering the research community's ability to evaluate these solutions or build upon them effectively.
\end{itemize}

\section{Concepts for Integrating Legacy Systems}
\label{sec:concepts}
In general, aggregating servers are designed to provide a centralised node for managing access from external clients~\cite{BMWi2019SichereOPCUA}. 
Legacy device integration typically lies outside the regular scope of aggregating servers and is usually realised via gateways connecting to the shop floor, e.g.,~\cite{etz2020}.
To analyse the different concepts for OPC UA-based aggregation in industrial networks, we first discuss the threat model that is considered (Section~\ref{sec:threatmodel}). 
Afterwards we discuss the identified concepts (C.1-3) for aggregation in Sections~\ref{sec:c1conduit} to \ref{sec:c3tcplevelaggregation}.

\subsection{Threat Model}\label{sec:threatmodel}
\begin{figure}[htb]
	\includegraphics[width=\linewidth]{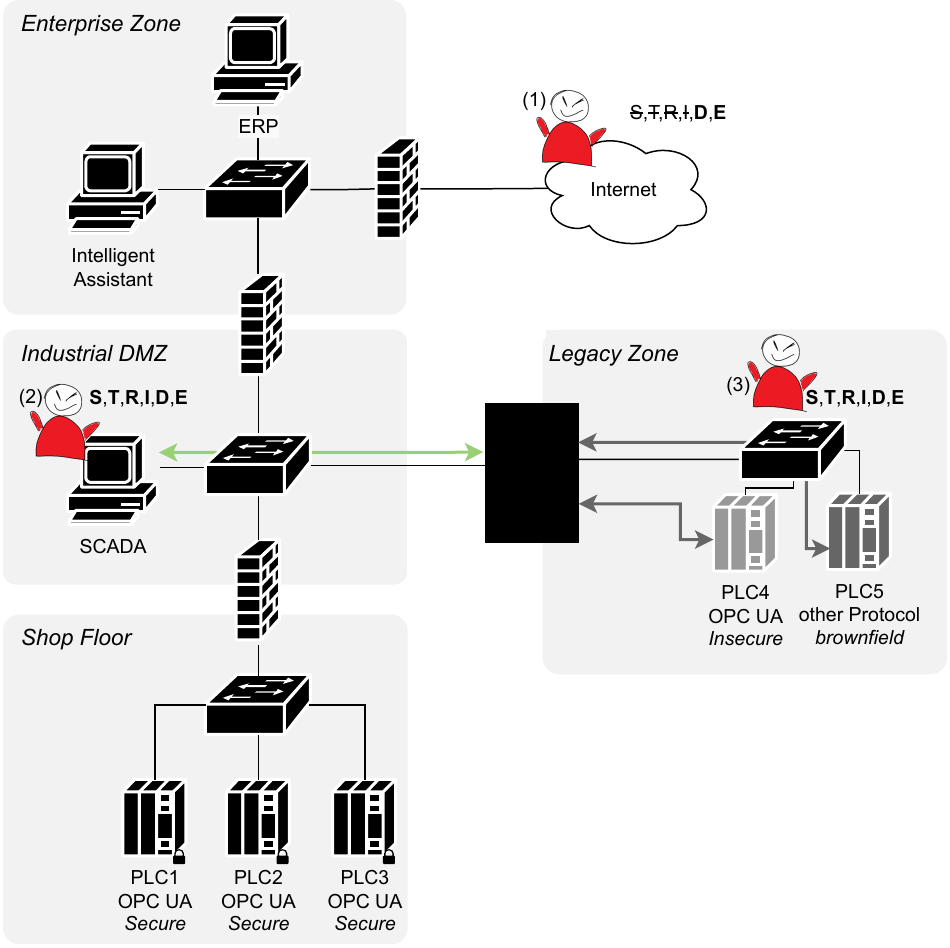}
	\caption{Threat model illustrating three different attacker levels while considering basic security controls enabled.}\label{fig:threat-model}
\end{figure}

Attacks are becoming increasingly sophisticated, as evidenced by the rising number of incidents in the OT domain~\cite{Anton2021GlobalState,fortinet2024ot}. 
Thus, we assume a skilled adversary at three different stages.
In the first stage, the attacker attempts to penetrate the network remotely via the internet. 
Next, the attacker gains a foothold in the industrial \gls{dmz}, and finally, the attacker compromises one of the legacy systems. 
Figure~\ref{fig:threat-model} illustrates an industrial automation network inspired by Purdue model and the location of the attacker, which is highlighted with (1) - (3).
The black box between the industrial \gls{dmz} and the legacy zone is a placeholder for the identified three concepts for aggregation (Sections~\ref{sec:c1conduit} to~\ref{sec:c3tcplevelaggregation}).

\textbf{Attacker Level 1.} The attacker is positioned outside the organisation and can perform \gls{dos} or privilege escalation attacks. 
Regarding the latter, an example may be that the attacker manages to exploit a vulnerability in a application, which allows them to get further access into the system (attack level~2).

\textbf{Attacker Level 2.} The attacker has gained access to a system located in the industrial \gls{dmz}, allowing them to execute \gls{dos} and privilege escalation attacks against the legacy zone. 
This scenario can be divided into two cases: (i) the attacker obtains physical access to the network through social engineering (e.g., by posing as an engineer or service staff) and compromises a device, or (ii) the attacker exploits a firewall vulnerability to gain network access. 
In both cases, the attacker can not only perform \gls{dos} and privilege escalation attacks, as in attacker level~1, but also conduct Spoofing, Tampering, Repudiation, and Information Disclosure attacks against the legacy zone

\textbf{Attacker Level 3.} Through social engineering or other actions (as demonstrated with Stuxnet~\cite{mueller2012stuxnet}), the attacker manages to access the legacy zone, hence they are able to perform attacks associated to all six STRIDE threats.

\subsection{Conduit to a legacy zone (C.1)}\label{sec:c1conduit}
In the context of integrating a legacy zone into an industrial network, one common approach is to use conduits, such as firewalls, to manage and secure access to and from the legacy devices.

By employing a firewall as a conduit, all data exchanged between the legacy zone and other network segments can be monitored, filtered, and restricted based on predefined security policies. This configuration minimises the risk of unauthorised access or malware propagation from the legacy devices to the broader network. The firewall can enforce rules such as allowing only specific protocols, limiting communication to essential systems, or implementing intrusion detection and prevention mechanisms.

However, this approach has some significant disadvantages. Clients from the ``secure'' zones, the industrial network, may still need to communicate using outdated and insecure protocols that are compatible with the legacy devices, exposing them to potential risks. 
Furthermore, deep packet inspection (DPI) would be required to enhance security by thoroughly analysing the data traffic for malicious content, requiring more resources.

\subsection{Application-Level Aggregating Server (C.2)}\label{sec:c2applicationlevelaggregation}
The second concept for connecting a legacy zone to an industrial network involves the aggregation on application level, which is commonly implemented using gateways/proxies or specialised solutions like OPC UA aggregating servers. 
These aggregating servers are specifically designed for scenarios where external entities need to access, both read and write, OPC UA nodes within the industrial automation network. 
 
This approach not only simplifies management and monitoring but also enhances interoperability by presenting data in a standardised format, reducing the need for direct interaction with insecure legacy protocols. 
However, the effectiveness of this scenario depends on the capabilities of the aggregation logic and the completeness of the device-specific data mapping.
Moreover, aggregating servers were neither designed to be (i) used for handling access to legacy systems within the network, and (ii) used as a security solution.

A challenge of this approach is \textit{namespace pollution}, as the namespace is no longer unique to individual devices. 
Instead, it becomes a shared space, which can lead to conflicts, ambiguities, or difficulty in managing and distinguishing data from different devices.

Graube et al.~\cite{graube2017} further detailed some disadvantages.
One notable issue mentioned is that industrial automation software is typically designed with a \textit{static, fixed namespace}, which relies on constant node IDs for access. 
The \textit{persistence} of these node IDs can pose a challenge as well, as changes to the merged information model may affect the order of nodes represented by the aggregating server during runtime.

Another challenge which needs to be taken into account is that an application-level aggregating server may introduce \textit{timing issues}.
Particularly in cases when \glspl{plc} are required to communicate with each other in real-time, potentially affecting the overall performance and responsiveness of the industrial system, which often relies on precise and synchronised operations.
	
\subsection{TCP-Level Aggregating Server (C.3)}\label{sec:c3tcplevelaggregation}
Individual namespaces for each device within the legacy zone can be maintained by distributing them at the transport layer (TCP) rather than at the application layer, as is done by \gls{opcua} aggregating servers (\textit{C.2}). 
This solution requires multiple \gls{opcua} servers running on the aggregating server, allowing devices to be distinguished through their respective TCP ports.

Timing constraints still need to be accounted for, as this solution does not eliminate all the previously mentioned disadvantages of \textit{C.2}.
However, it effectively addresses issues such as \textit{namespace pollution} and the reliance of industrial automation software on a \textit{static, fixed namespace}. 

Challenges arise from the dynamic deployment of multiple \gls{opcua} servers.
One such challenge is the need to map the TCP ports to devices in the legacy zone using static mappings.
These static mappings are necessary to avoid re-configuring \glspl{plc} whenever changes occur within the legacy zone.
Once the ports are configured, however, the \glspl{plc} do not require further adjustments.

Another challenge stems from the use of separate TCP ports for each legacy device. 
Specifically, integrating new PLCs -- particularly a large number of them -- may be more time-consuming and requires also more adjustments to the firewall compared to the approach described in \textit{C.2}.

As for any other concept presented for our use case, a TCP-level aggregating server must reflect the current status of the legacy devices requiring it to communicate via the legacy protocol. 
The monitoring of the operational state of these devices is necessary, due to the otherwise high time delay for accessing the \gls{opcua} server representing the legacy device.

\subsection{Additional Security Measures}
Regardless of the discussed concepts OPC~UA must follow state-of-the-art security measures when used in production. 
Moreover, the legacy zone, as an insecure network, must be additionally protected.
Network filters and deep packet inspection may be required, especially when considering attacker level 3 in the threat model (see~\cref{sec:threatmodel}).

Additional measures to secure the vulnerable legacy zone include the deployment of honeypots.
For instance, a network filter facing the secure zone can redirect suspicious traffic to a sandboxed honeypot \cite{SSRH24}.
The advantage of deploying a honeypot is that it allows one to immediately detect, record and learn from the attacker's behaviour.

\section{\sigmaserv}
\label{sec:sigmaserv}
The architecture of~\sigmaserv{} builds upon the TCP-level aggregating server design (C.3) presented in~\cref{sec:c3tcplevelaggregation},
as it mitigates \textit{namespace pollution} and minimises modifications to existing industrial software relying on a fixed namespace~\cite{graube2017}.

\subsection{Architecture}\label{sec:architecture}
\begin{figure}
	\centering
	\includegraphics[width=1\linewidth]{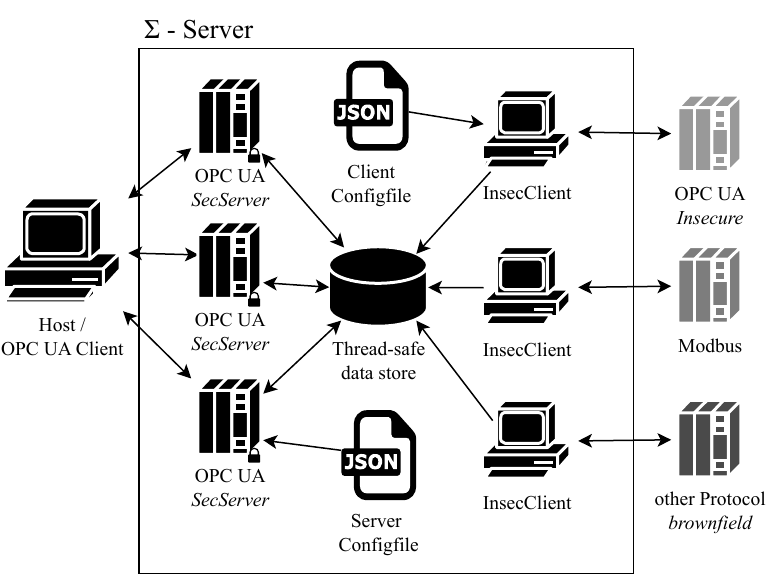}
	\caption{Architecture of the proposed \sigmaserv{} illustrating the separation of insecure client interfaces and multiple secure \gls{opcua} endpoints.}\label{fig:sigma-server-architecture}
\end{figure}

The \sigmaserv~demonstrates a proof-of-principle implementation 
and is implemented with the open62541\footnote{\url{https://github.com/open62541/open62541/}} (v.1.4.8) library.
Its modular design is based on three main components: multiple insecure clients (\textit{InsecClients}), a central thread-safe data store, and multiple secure servers (\textit{SecServers}).
The overall architecture is shown in~\cref{fig:sigma-server-architecture}.

Each InsecClient is responsible for a single legacy device, such as a PLC or another system that relies on insecure or outdated communication protocols.
The InsecClients operate in separate threads and establish connections to the assigned legacy endpoints.
They retrieve both structural information and real-time data values, which are stored in the central thread-safe data store.
The current version of \sigmaserv{} supports \gls{opcua} and Modbus/TCP protocols. However, the architecture allows the integration of additional insecure communication protocols, such as CAN or PROFINET, by adding further InsecClient variants in the publicly available open-source codebase.

The shared storage functions as the central data buffer of the system, maintaining two synchronised thread-safe maps: one for the static node structure and another for current data values.
This design enables strict separation between the insecure and secure zones.
All structures and data are mirrored to JSON files under \textit{config/<alias>/namespace<N>.json} allowing independent inspection of the running system.

The SecServers host a secure \gls{opcua} server instance per legacy device.
Each SecServer runs on a dedicated TCP port (e.g., 4841) and provides a unique \gls{opcua} endpoint to represent a single device from the insecure network.
This approach guarantees full namespace separation, thereby eliminating the problem of namespace pollution and ensuring a clear one to one mapping between legacy devices and their secure OPC UA representations.
Thus, hosts in the secure zone can access legacy devices by reconfiguring only the TCP port used for communication.
Communication between clients in the secure zone and the \sigmaserv~is encrypted and authenticated using \gls{opcua} SecurityPolicy \textit{Basic256Sha256}, supporting both certificate-based and username/password authentication.

At system startup, the \sigmaserv{} loads two configuration files: the client configuration defines all insecure endpoints, node IDs, and Modbus registers to be queried.
The server configuration specifies which secure servers to start, along with their corresponding aliases and TCP ports.
After initialisation, each InsecClient retrieves the node structures from its source device and files the shared storage with values.
The SecServers then register these structures into their local \gls{opcua} address space and starts providing real-time values to the host systems such as \gls{scada}.

\begin{figure}
	\centering
	\includegraphics[width=1\linewidth]{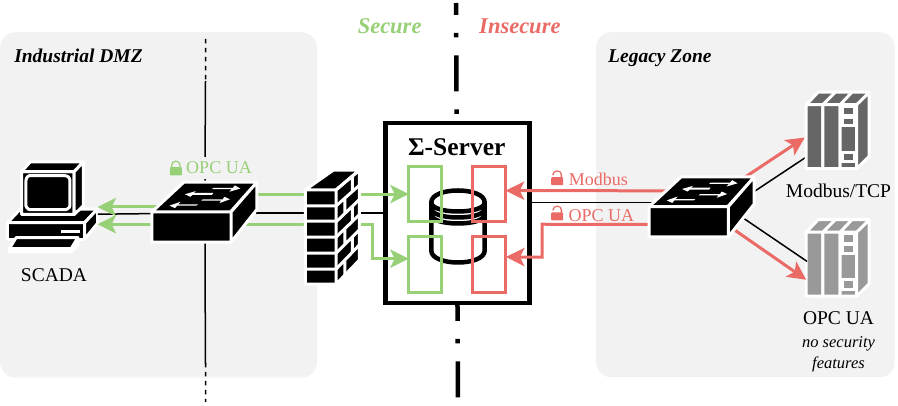}
	\caption{Utilisation of \sigmaserv{} as a bridge between the secure zone and the insecure legacy zone. }\label{fig:sigma-server-zones}
\end{figure}

\cref{fig:sigma-server-zones},  shows the positioning of the TCP-level aggregating server, acting as a bridge between secure and insecure network zones. Both network zones are connected to a network interface of the \sigmaserv{}.

\subsection{Operation Principle}

When the \sigmaserv{} starts, the main process loads the configuration files and creates a separate thread for each client and server defined in the setup.
Each InsecClient thread connects to its assigned device. For OPC UA devices, the client first reads the NamespaceArray (standard node i=2255) to map each namespace index to its URI.
This information is stored in memory as \texttt{<Alias>:<NsIndex> $\to$ <URI>} and later used by the secure servers to reconstruct the original namespace table.
The client subsequently reads the configured node attributes such as \textit{DisplayName}, \textit{Description}, \textit{DataType} and \textit{BrowseName}.
These attributes are queried using the open62541 read service and saved together with the nodes' numeric identifiers (\textit{NamespaceIndex}, \textit{NodeId}).
The complete browse path is constructed by recursively traversing parent-child relationships up to the root folder.
Each path is prefixed with the device alias (for example Objects/PLC21/...) so that every device has its own subtree in the global structure. 

This structure is stored in a thread-safe map and also written to JSON files under \textit{config/<Alias>/Namespace<N>.json}.
These files contain a minimal list of all node identifiers for the given namespace and allow dynamical adjustments to the currently needed nodes.
Once the reading the structure is finished, the client switches into a cyclic read loop.
It continuously reads the values of all configured nodes or, in the case of Modbus devices, the defined registers.
The values are normalised into basic OPC UA types: Boolean, Int16/32/64, Float, Double, String, or DateTime.
For timestamps, the current system time is stored as an OPC UA DateTime value or as Unix time, depending on the source.
Modbus registers are converted according to the configuration, for example scaling a raw integer by $0.1$ to represent a temperature value.
The resulting data is written atomically to the global value map, replacing the previous entry.
Each entry is identified by a deterministic key of the form \texttt{<Alias>:<NamespaceIndex>:<NodeId>}.

The shared storage acts as the only communication interface between insecure clients and secure servers.
It is implemented as a thread-safe key-value map protected by internal mutexes.
This ensures consistent access even when multiple threads write simultaneously.
Clients only write new values, while the secure servers only read them. 
No direct network communication exists between these two domains, which ensures full isolation between the insecure and secure network zones. 
The current implementation focuses exclusively on read operations from legacy devices to secure clients.

On the secure side, each SecServer thread starts once the structure data is available in memory.
During its initialisation, it reads all stored namespaces for its assigned alias and adds them to its own OPC UA server configuration.
Each variable node is created with the same numeric NodeId(ns, id) that was read from the original device.
Folder hierarchies are recreated from the stored browse paths.
A read callback is registered for every variable.
When a secure OPC UA client requests a value, the callback retrieves the corresponding JSON object from the shared data map, converts the stored value to the correct data type based on the saved DataTypeId, and returns it as a UA\_Variant.
This conversion supports all common scalar types and ensures that data are presented exactly as they were read from the original device.
The server's namespace table is rebuilt from the stored URIs, and the OPC Foundation standard namespace (\textit{http://opcfoundation.org/UA/}) is not mirrored to avoid collisions.

Each client and server operates in its own thread and is designed with a stoppable control mechanism that enables proper shutdown.
In the case of a connection loss or device failure, the affected client continuously attempts to re-establish communication with its assigned endpoint. 
The reconnection mechanism is handled within the regular polling loop and does not require manual actions. 

In addition to the conventional client–server mode, the implementation also supports a publish–subscribe mode.
In this configuration, the secure OPC UA servers act as publishers, periodically transmitting aggregated data to one or more subscribers within the secure zone. The operational mode can be switched between the two modes directly through the server configuration file \textit{server\_configuration.json}.

The implemented \sigmaserv~is fully platform independent, which is achieved through containerisation using Docker.
This approach allows the entire architecture to be deployed on both Windows and Linux hosts without any code modifications.

\section{Experimental Setup}\label{sec:experimental:setup}

\subsection{Test Environment}\label{subsec:Test Environment}

\begin{figure}
	\centering
	\input{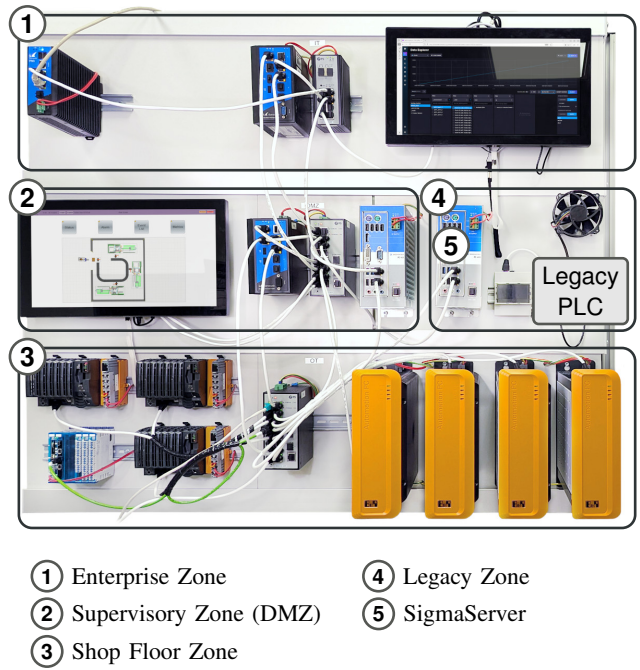}
	\caption{The testbed setup for the deployment is segmented into four zones, with the insecure legacy zone being integrated into the DMZ through the proposed \sigmaserv.}\label{fig:testbed}
\end{figure}

To evaluate the operational behaviour and performance of the \sigmaserv, the legacy zone implemented in the testbed of the \gls{jrc} \gls{isia} (see \cref{subsec:ref-arch}) was utilised.
The setup reflects a realistic replicable industrial environment, allowing data flow, latency and potential attack vectors to be evaluated under controlled conditions.
The focus of the experiment lies on the connection between the secure network zone and the legacy zone which is connected through the \sigmaserv{}.
The physical setup is illustrated in \cref{fig:testbed}.

The \sigmaserv{} is deployed on a Sigmatek IPC~511 running Windows 10 IoT Enterprise.
This IPC is equipped with an Intel Celeron G1820 CPU, 8 GB RAM and two Ethernet interfaces.
The interfaces are configured to provide the connection to the secure and legacy networks respectively.
This hardware was selected to reflect a typical industrial production environment.

In the legacy zone, two legacy device types are used.
One being a B\&R X20CP1686X PLC acting as the injection-molding machine.
For this experiment, this PLC is configured with an insecure OPC UA interface and access to multiple nodes.
The PLC is polled by an InsecClient which first retrieves the namespace array and node attributes, then periodically reads configured values.
The other legacy device is a Raspberry Pi 4 acting as a Modbus/TCP server representing a cooling system. 
The Modbus server is implemented using the pymodbus library\footnote{\url{https://pymodbus.readthedocs.io/en/latest/}} and provides two holding register containing: the temperature in degree Celsius~(°C) from a DS18B20 sensor, and fan speed in revolutions per minute~(RPM).
The \sigmaserv's Modbus InsecClient polls these registers at a configurable interval and writes normalised values into the central thread-safe data store.

On the secure side, two SecServer instances mirror the collected data and present it via secure \gls{opcua} endpoints (TCP ports 4841 and 4842).
A host system, in this experiment UaExpert, connects to these endpoints using the security policy \textit{Basic256Sha256} with certificate-based authentication enabled.

The network topology is realised using an unmanaged switch in the legacy zone connecting the OPC UA insecure and Modbus server to the \sigmaserv{}. 
A Barracuda F183R firewall was used to enforce strict TCP routing between the zones.
Therefore, only required ports are allowed (i.e., TCP port 4841-4842).

For the evaluation described in~\cref{subsec:Performance Metric}, an external test client was used, running on an HP Z2 Tower G9 Workstation (Intel Core i7-12700, 32GB RAM).
The client operated on a Linux system with a generic kernel (6.14.0-generic).

It is important to note that neither the \sigmaserv{} Windows 10 IoT PC, nor the test client utilised a hard real-time operating system kernel.
This setup was chosen to represent a typical industrial monitoring scenario where servers and clients operate on typical IT infrastructure.

\subsection{Attack Scenario}\label{subsec:Attack Scenario}

Based on the threat model introduced in \cref{sec:threatmodel}, the following attack scenario illustrates the current security posture of the \sigmaserv.
The implemented architecture effectively mitigates attacks associated with attacker level~1 and level~2 adversaries.
External attackers or compromised systems within the industrial \gls{dmz} cannot perform spoofing, tampering, repudiation or information disclosure attacks.
Although the \sigmaserv{} provides strong isolation through strict network segmentation as well as authenticated and encrypted communication, it cannot fully prevent \gls{dos} or privilege escalation attempts.
Its multi-port design inherently distributes the processing load across independent secure server instances, limiting the potential impact of targeted \gls{dos} attacks to individual endpoints rather than the entire system.
Comprehensive defense against such attacks would still require additional protective measures.

The level 3 attacker, who gains access to the legacy zone itself, remains a realistic and critical threat.
In the considered attack scenario, the adversary infiltrates the production network via social engineering by impersonating a maintenance technician.
Network access is obtained by connecting a computing device like a Raspberry Pi Zero to an unused Ethernet port on a switch linking critical production systems.
The legacy zone includes both an insecure OPC UA server running on a \gls{plc} representing an injection molding machine, and a Modbus/TCP cooling system as described in~\cref{subsec:Test Environment}.
Subsequently, the attack focuses on the Modbus protocol.

Using this position, the attacker carries out a spoofing attack, more specifically, a targeted ARP poisoning attack on the unsecured communication channel between the aggregation server and the Modbus server.
This attack enables the adversary to manipulate the data transmitted from the Modbus server to the aggregation server.
By utilising Ettercap and custom Etterfilter rules, the TCP payload of Modbus responses is continuously modified during transmission, causing the aggregation server to provide falsified data to the user.

For example, the attacker can present operators with stable and risk-free temperature readings from the cooling system, while in reality the device is overheating due to the insufficient thermal management.
As a result, workers remain unaware of the serious condition, potentially leading to severe damage.
In a real world production environment, this could result into equipment failure, production downtime, and therefore significant financial losses.

Besides the demonstrated ARP spoofing attack, the analysis highlights that the attack surface of insecure fieldbus protocols is diverse. Replay attacks, denial-of-service attempts on the Modbus port, or the deployment of rogue Modbus servers represent further realistic threat scenarios.
In industrial environments, this does not only pose a security risk but also directly impacts safety, as manipulated process values can cause physical damage to equipment, production lines and endanger human life.

The presented proof-of-principle of \sigmaserv{} cannot mitigate these weaknesses on its own but must be complemented by additional mechanisms such as \glspl{ids}.
While the design and evaluation of such countermeasures are beyond the scope of this article, they represent a promising direction for future research.  

\subsection{Evaluation Metrics}\label{subsec:Performance Metric}

To evaluate the operational efficiency of the \sigmaserv, three experiments were conducted focusing on (i)~end-to-end latency between the insecure and secure interfaces under realistic operating conditions (ii)~internal software latency of the \sigmaserv{} between InsecClient and SecServer and (iii)~average CPU and RAM utilisation compared to the OPC Foundation Console Aggregation Server for different aggregation scenarios.

\begin{figure}
	\includegraphics[width=\linewidth]{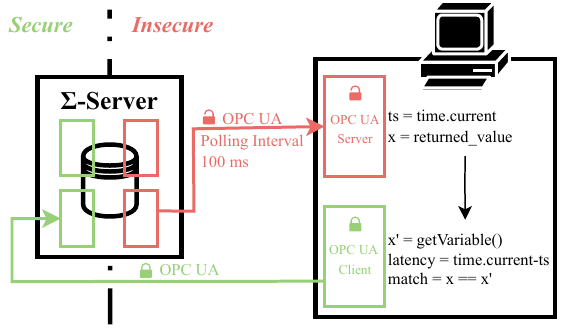}
	\caption{Testing setup for measuring the latency of \sigmaserv{} and other solutions.}
	\label{fig:latency_testing_setup}
\end{figure}

{\bf End-to-end \sigmaserv~latency}:
The first experiment evaluates the end-to-end latency of the \sigmaserv{} using the testing setup shown in \cref{fig:latency_testing_setup}.
This measurement includes the internal processing time of the \sigmaserv{} and the network latency between the participating systems.
As illustrated in \cref{alg:latency_measurement}, the external latency test client generates a random value each time it is polled by the 
\sigmaserv{}'s InsecClient.
Immediately after this value is returned, the SecTestClient reads the corresponding value from the secure OPC UA endpoint of the \sigmaserv{}.
The latency is defined as the time difference between the moment a value is generated at the InsecTestServer and the moment the same value is successfully retrieved from the SecServer.
This reflects a realistic scenario where data from an insecure legacy PLC flows through the \sigmaserv{} to a secure OPC UA client.

To verify whether the values are matching, the test compares the received value from the secure interface with the returned value from the insecure interface.
If both values match, the result is marked as \textit{true}, otherwise, it is marked as \textit{false}.
This true/false validation ensures that only matching values are considered in the evaluation but also shows if the test-client is fast enough for a proper validation.

Time measurements for the end-to-end latency were performed using Python's \texttt{time.monotonic()}, which utilises the underlying Linux CLOCK\_MONOTONIC to ensure high-resolution monotonic timestamps.

\begin{algorithm}
	\caption{End-to-end latency measurement}
	\label{alg:latency_measurement}
	\begin{algorithmic}
		\State \textbf{Participants:} InsecTestServer, SecTestClient, InsecClient, SecServer
		\State \textbf{Note:} InsecTestServer and SecTestClient reside on the same external test device
		\State
		\State InsecTestServer generates random value when read by \sigmaserv{}
		\State $returnedValue \gets \text{InsecTestServer.getGeneratedValue()}$
		\State $ts \gets \text{current\_time()}$
		\State $receivedValue \gets \text{SecTestClient.readFromSecServer()}$
		\State $latency \gets \text{current\_time()} - ts$
		\If{$returnedValue = receivedValue$}
		\State $match \gets \text{true}$
		\Else
		\State $match \gets \text{false}$
		\EndIf
		\State \text{save(latency, match, returnedValue, receivedValue)}
	\end{algorithmic}
\end{algorithm}

{\bf Internal software \sigmaserv{} latency}:
The second experiment focuses on the internal software latency of the \sigmaserv.
For this purpose, the implementation provides built-in software latency measurements around the shared data store.
Two timestamps are recorded: 
\begin{itemize}
	\item \textit{$\Delta{t_1}$} measures the time between the arrival of a new value at the InsecClient and the completed write operation into the thread-safe data store.
	\item \textit{$\Delta{t_2}$} measures the time between the SecServer receiving a read request from a secure OPC UA client and the moment the requested value is fetched from the data store and is available on the SecServer for transmission.
	\item \textit{$T_{proc}$} is therefore defined as the overall internal software latency, $\Delta{t_1}+\Delta{t_2}$.
\end{itemize}

These measurements capture the internal processing behaviour of the \sigmaserv{} between the insecure and secure sides, while explicitly excluding any polling interval, as well as any network latency. The internal software latency \textit{$T_{proc}$} was measured using the C++~\texttt{std::chrono::high\_resolution\_clock}, configured to record timestamps with microsecond resolution.
Cryptographic operations required for secure OPC UA communication are not included in the internal software latency measurement.

\textbf{Server resource utilisation:}
The third experiment evaluates the runtime efficiency of \sigmaserv{} and compares it to the OPC Foundation Console Aggregation Server. 
To analyse scalability and resource behaviour under different workloads, four test scenarios were conducted, ranging from a single active OPC UA server to multiple concurrent OPC UA servers. Specifically, the measurements included configurations with one, two, and three simultaneously running OPC UA servers, as well as a fourth setup combining three OPC UA servers with one Modbus client to evaluate the mixed-protocol performance of the \sigmaserv. Since the fourth setup requires Modbus handling, it can only be run on the \sigmaserv. 

All experiments were performed on the JRC ISIA testbed (see Section~\ref{subsec:ref-arch}) using the configuration described in \cref{subsec:Test Environment}.

\section{Results \& Discussion}\label{sec:results:discussion}

The performance evaluation of the \sigmaserv~was carried out according to the methodology defined in \cref{subsec:Performance Metric}.
All data were recorded under the same conditions.

\subsection{End-to-End \sigmaserv~Latency}

A total of 14,287 samples were recorded for this experiment, with every single measurement resulting in a \textit{true} match between the value generated on the insecure interface and the value read from the secure OPC UA endpoint.
The measured latency values range from a minimum of 2.6 ms to a maximum of 24.7 ms with a standard deviation of 3.0 ms.

\begin{figure}
	\centering
	\input{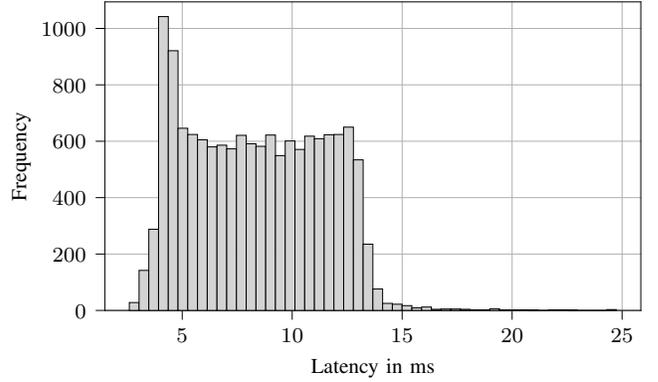}
	\caption{Distribution of the end-to-end \sigmaserv{} latency measured with the test client between the insecure and secure interfaces.}
	\label{fig:latency_sigmaserv}
\end{figure}

Since no mismatches occurred, the results indicate that the \sigmaserv{} consistently processes and forwards values faster than the external latency test client can resolve.
Although the exact minimum latency of the \sigmaserv{} cannot be determined due to the measurement limits of the test client, the results show that the complete data path, including the network transmission, is less than 2.6 ms.
\Cref{fig:latency_sigmaserv} shows the latency distribution of this experiment.

\subsection{Internal Software \sigmaserv~Latency}

\begin{figure}
	\centering
	\input{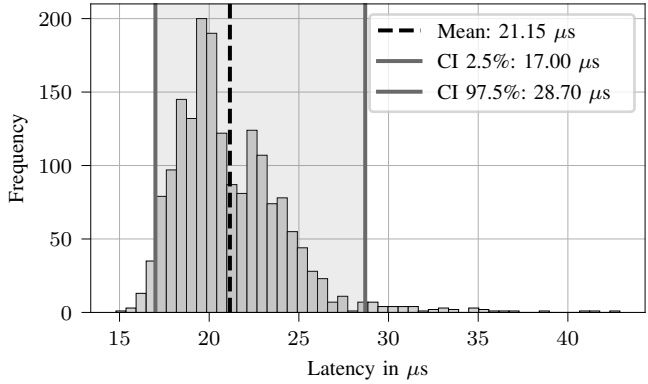}
	\caption{Internal software latency of the \sigmaserv, showing the distribution of processing delay between InsecClient and SecServer.}
	\label{fig:sigma_ptime_boxplot}
\end{figure}

A total of 1,788 samples were collected to determine the internal software delay of the \sigmaserv.
The measured latencies range from a minimum of 14.80 $\mu$s to a maximum of 42.90 $\mu$s with a standard deviation of 3.11 $\mu$s. The overall internal software latency \textit{$T_{proc}$} was measured at 21.15 $\mu$s. \Cref{fig:sigma_ptime_boxplot} summarises the latency distribution and shows the corresponding 95\% confidence interval.
As this measurement excludes polling intervals, network latency and cryptographic operations, it isolates the pure computational overhead of the thread-safe data store and the data transfer between InsecClient and SecServer.

\subsection{Server Resource Utilisation}

CPU and RAM utilisation were recorded over a one-hour period using the \texttt{docker stats} monitoring tool, with sampling intervals of one second. The polling interval of the InsecClient was configured to 100 ms.

The average CPU load of the \sigmaserv{} increased nearly linearly with the number of concurrently running server instances, ranging from 0.75\% for a single OPC UA server to 3.16\% when three OPC UA servers and one Modbus client were active.
The RAM usage showed similar predictable behaviour, increasing from 6.12 MiB at lowest to 18.79 MiB at maximum.

\begin{figure}
	\centering
	\input{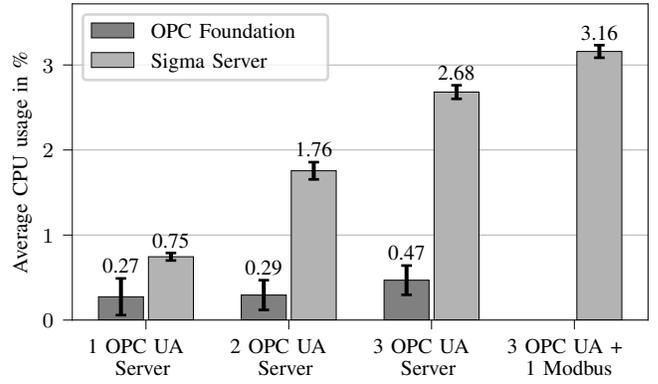}
	\caption{Average CPU usage of the \sigmaserv{} compared to the OPC Foundation Console Aggregation Server.}
	\label{fig:sigma_cpu_bar}
\end{figure}

\begin{figure}
	\centering
	\input{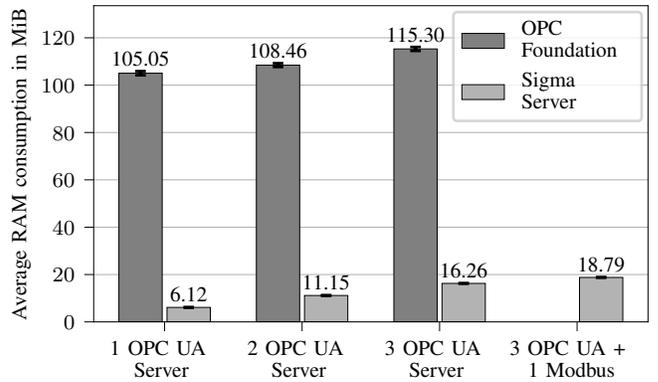}
	\caption{Average memory consumption of the \sigmaserv{} compared to the OPC Foundation Console Aggregation Server.}
	\label{fig:sigma_mem_bar}
\end{figure}

\Cref{fig:sigma_cpu_bar} presents the average CPU utilisation for all scenarios, while \cref{fig:sigma_mem_bar} compares the RAM utilisation of the \sigmaserv{} with the OPC Foundation Console Aggregation Server.
Both implementations were configured with an identical OPC UA address space to ensure a fair comparison.
As noted earlier, the mixed protocol scenario (OPC UA and Modbus) is only possible with \sigmaserv{}, hence \Cref{fig:sigma_cpu_bar} only includes the results for our solution.

\subsection{Discussion}

The evaluation demonstrates that the \sigmaserv~requires only minimal computing resources while maintaining stable and predictable behaviour in all scenarios examined.
The overall internal software latency, measured in the low microsecond range, confirms that the thread-safe data store and the modular client–server design operate efficiently and no significant delays occur.

The resource utilisation results highlight clear architectural differences compared to the OPC Foundation Console Aggregation Server.
\sigmaserv{} uses multiple independent OPC UA server instances with TCP-level separation, while the OPC Foundation Console Aggregation Server provides a single, centrally aggregated address space.
As a results, \sigmaserv{} shows slightly higher CPU usage, it remains within a low and predictable single-digit range and scales linearly with the number of aggregated devices.
At the same time, \sigmaserv{} requires significantly less RAM and shows a noticeably lower variance in both CPU and RAM usage.

Overall, the results confirm that the \sigmaserv~offers a reliable and resource efficient mechanism for bridging insecure field devices with secure OPC UA infrastructures, while keeping software overhead minimal and ensuring the elimination of namespace pollution.
Together with the predictable performance characteristics, the results confirm that the \sigmaserv~meets the practical requirements of brownfield integration by providing secure OPC UA endpoints with low overhead.

\begin{figure}
	\centering
\def\participantNames{Insecure Server, SigmaClient, SigmaServer, Secure Client}
\def\timeMax{12}    
\def\yStep{1.4}    
\def\xStart{1}     
\def\xScale{0.5}  
\def\labelOffset{0.5}
\def\pollingOffset{3}

\newcommand{\horizontaldim}[5]{%
	\path (#2);                    
    \pgfgetlastxy{\tempX}{\ycoordB}  
	\path (#3);                    
    \pgfgetlastxy{\tempX}{\ycoordA} 
	\pgfmathsetmacro{\Ymax}{max(\ycoordA,\ycoordB)}
	\pgfmathsetmacro{\Ymin}{min(\ycoordA,\ycoordB)}
    \dimline[%
        extension start length=#1,
        extension end length=\Ymax -(\Ymin)  + #1,
        label style={font=\footnotesize, #5, fill=none, draw=none}
    ]
    {($(#2)+(0,#1)$)}{($(#3 |- #2)+(0,#1)$)}{#4}%
}

\begin{tikzpicture}[
    event/.style={rounded rectangle, draw , fill=gray!10, minimum height=4pt, inner sep=0pt, outer sep=0pt,},
    timeline/.style={thick, -Latex},
    arrow/.style={-Latex},
    font=\footnotesize
]

\begin{scope}[xscale=\xScale, yscale=-1/\yStep, shift={(\xStart, 0)}, transform shape=false]
	\coordinate (topLeft) at (-\labelOffset-4.2, 1-0.4); 
	\coordinate (bottomRight) at (\xStart+\timeMax+0.2, 2.4);
	
	\draw[ fill=gray!20, rounded corners=5, draw=none] (topLeft) rectangle (bottomRight);

	\node[anchor=north] at (\xStart+\timeMax, 3.2)  {t};
	
	\newlength{\nodewidth}
	\pgfextractx{\nodewidth}{\pgfpointxy{\xScale}{0}} 
		\foreach \name [count=\i] in \participantNames {
			\pgfmathsetmacro{\y}{(\i-1)}
	
			\node[anchor=east] at (-\labelOffset, \y) {\name};
	
			\draw[timeline] (0, \y) -> (\xStart+\timeMax, \y);
	
			\foreach \t in {0,1,...,\timeMax} {
			    \draw (\t, \y+0.2) -- (\t, \y-0.2);
			}
		}
	\foreach \offset [count=\i] in {0,7} {
		\coordinate (delayStart\i) at (\offset+2, 0);
		\coordinate (delayEnd\i) at (\offset+\pollingOffset+2, 3);
		\draw[fill=gray, fill opacity=0.3, draw=none] (delayStart\i) rectangle (delayEnd\i);
	
		\node[event, minimum width=2\nodewidth, anchor=west] (Poll1\i) at (\offset, 1 ) {}; 
		\node[event] (Process1\i) at (\offset+1, 0 ) {}; 
	
		\node[event, minimum width=2\nodewidth, anchor=west] (Poll2\i) at (\offset+\pollingOffset, 3 ) {}; 
		\node[event] (Process2\i) at (\offset+\pollingOffset+1, 2 ) {}; 
	
		\draw[arrow] (Poll1\i.north west) -- (Process1\i) node[midway, left] {};
		\draw[arrow] (Process1\i) -- (Poll1\i.north east) node[midway, left] {};
	
		\draw[arrow] (Poll2\i.north west) -- (Process2\i) node[midway, left] {};
		\draw[arrow] (Process2\i) -- (Poll2\i.north east) node[midway, left] {};
		
	}
\end{scope}

\horizontaldim{0.2cm}{Process11}{Process12}{Polling Interval $T$}{above};
\horizontaldim{0.2cm}{delayStart2}{Poll22.east}{Forwarding Delay $t_d$}{above};
\horizontaldim{-0.2cm}{Poll12.west}{Poll12.east}{Transmission Time $2 \cdot t_t$ }{below};
\horizontaldim{-0.2cm}{Poll21.west}{Poll11.west}{ Shift $t_s$}{below};

\end{tikzpicture}
\caption{Timing diagram illustrating the realistic forwarding delay $t_d$ introduced by the \sigmaserv. 
	A common polling interval of $T$ is assumed, with each polling operation requiring $2 t_t$. 
	The polling activities are shifted by $t_s$, resulting in a forwarding delay 
	$t_d \in [t_t,\, t_t + T)$.}
	\label{fig:forwarding_delay}
\end{figure}
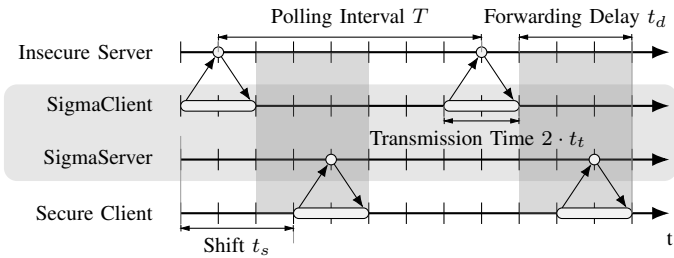

While the performed latency measurements show that the \sigmaserv{} introduces only minimal additional latency, 
we also need to consider typical deployment scenarios. 
In these scenarios the resulting latency still can be significant, yet they primarily depend on the timing of the polling or publish-subscriber mechanisms employed by the clients.

To illustrate this behavior in more detail, we discuss the forwarding latency that may appear when both the \sigmaserv{} client and the secure client use polling to retrieve data updates. 
\Cref{fig:forwarding_delay} shows a timing diagram of this scenario.
For this analysis, we assume a common polling interval $T$ for both participants.
The processing time of each participant is neglected, as it is at least an order of magnitude smaller than the polling interval $T$. 
Similarly, any internal data sampling performed on the insecure server is ignored, since it occurs regardless of whether the \sigmaserv{} is present.
The expected additional forwarding delay for new data arriving via the \sigmaserv{} depends on the time shift $t_s$ between the two polling activities. 
As shown in \cref{fig:forwarding_delay}, this delay lies within the interval $t_d \in [t_t,\, t_t + T)$,
where the lower bound represents the best case and the upper bound (excluded) corresponds to the worst-case delay. 

When considering the overall data age, with a client directly polling the insecure server at interval $T$, this results in a maximum data age of $t_t + T$, 
which is the sum of the transmission time and the polling interval.
Introducing the \sigmaserv{} in between, with an equal polling rate $T$, leads to a best-case data age of $2 \cdot t_t + T$ and a worst-case data age of $2 \cdot (t_t + T)$. 

\section{Conclusion}\label{sec:conclusion}
In this paper, we address the challenge of securely integrating insecure legacy devices into OPC UA based industrial automation networks.
While OPC UA aggregating servers and brownfield retrofitting solutions have been widely discussed in prior work, security aspects are rarely considered.
To close this gap, we analysed existing aggregation and retrofitting approaches from a security perspective. 
Furthermore, we developed a threat model for a typical industrial network setup to identify potential attack vectors.
Based on these analyses, three architectural concepts for bridging secure and insecure network zones were derived and evaluated.

We identified TCP-level aggregation as the most suitable concept for the secure integration of legacy systems. 
It prevents namespace pollution and establishes a clear one-to-one mapping between legacy devices and their secure representations, minimising the integration effort required.
Based on this concept, we introduced \sigmaserv{} as a proof-of-principle implementation.
Its architecture enforces strict separation between insecure client interfaces and secure OPC UA endpoints by employing a central thread-safe data store and multiple secure servers listening on different ports.

The experimental evaluation on an industrial testbed demonstrated that \sigmaserv{} introduces negligible overhead.
End-to-end latency remained below 2.6 ms, with an average internal processing delay of 21.15 $\mu$s.
Resource utilisation is compared to the performance of the OPC Foundation Console Aggregation Server, which focuses exclusively on \acrshort{opcua}.
In contrast, \sigmaserv{} supports bridging not only \acrshort{opcua} but also insecure legacy protocols such as Modbus.
The comparison further revealed that \sigmaserv{} consumes less RAM than the OPC Foundation Console Aggregation while maintaining predictable CPU usage.
These findings confirm that secure legacy system integration without namespace pollution can be achieved with minimal technical complexity and without compromising performance.

Overall, \sigmaserv{} offers a practical, security-focused solution for bridging insecure legacy devices into modern secure industrial networks.
By enabling secure OPC UA communication and efficient runtime performance, it demonstrates that TCP-level aggregation can be effectively realised.

\subsection{Future Work}\label{subsec:future work}
While the presented \sigmaserv{} already provides a robust foundation for secure data bridging, several extensions are planned for future work.
The protection of the legacy zone itself remains an open challenge.
As demonstrated in~\cref{subsec:Attack Scenario}, the aggregation server cannot prevent attacks that originate within the insecure network.
Future work will therefore focus on additional security mechanisms such as \acrlongpl{ids} and the integration of honeypots.
These measures aim to detect and analyse malicious activities and to direct suspicious traffic away from critical legacy devices (sandboxing).

Additionally, the current implementation focuses on read access to legacy devices.
An extension of \sigmaserv{} can be to support write access and method calls.
This includes the secure forwarding of OPC UA write requests and method calls from secure clients to legacy systems.

Finally, future work will further focus on the behaviour discussed in~\Cref{fig:forwarding_delay}, which highlights the forwarding delay introduced by asynchronous polling between insecure and secure interfaces.
While the internal processing latency of the \sigmaserv{} is negligible, the effective latency strongly depends on the interaction between polling activities.
This latency however could be minimized when using a publish-subscribe pattern on the secure side 
and synchronizing the publishing activities with the polling activities on the insecure side. 
Implementing such a synchronization mechanism would nearly remove the remaining performance-related drawbacks of \sigmaserv{}.

\section*{Acknowledgement} 

The financial support by the Austrian Federal Ministry of Economy, Energy and Tourism, the National Foundation for Research, Technology and Development and the Christian Doppler Research Association is gratefully acknowledged. 

\medskip
\renewcommand*{\bibfont}{\footnotesize}
\printbibliography

\end{document}
\typeout{get arXiv to do 4 passes: Label(s) may have changed. Rerun}